\documentclass[12pt, twoside]{article}
\usepackage[semicolon,round,sort&compress]{natbib}  %
\usepackage{hyperref}
\usepackage{amsmath,amsthm, amssymb, amsfonts}
\usepackage{booktabs}
\usepackage{pdflscape}
\usepackage{calc}
\usepackage{float}
\usepackage{framed}
\usepackage{indentfirst}
\usepackage{chronology}
\usepackage[toc,page]{appendix}
\usepackage{tablefootnote}
\usepackage{subfigure}
\graphicspath{{images/}}
\usepackage{caption}
\usepackage[a4paper,width=150mm,top=25mm,bottom=25mm,bindingoffset=6mm]{geometry} % this was incomplete
\usepackage{filecontents}
\usepackage{algorithm}
\usepackage{algorithmic}

\usepackage{graphicx}
\usepackage{verbatim}
\usepackage{multirow}
\usepackage{bbm}
\usepackage[utf8]{inputenc}
\usepackage[english]{babel}
\usepackage{mathtools}
\usepackage{natbib}
\usepackage{mathtools}

\usepackage{multicol}
\renewcommand{\bibpreamble}{\begin{multicols}{2}}
\renewcommand{\bibpostamble}{\end{multicols}}

\makeatletter
\newcommand{\distas}[1]{\mathbin{\overset{#1}{\kern\z@\sim}}}%
\newsavebox{\mybox}\newsavebox{\mysim}
\newcommand{\distras}[1]{%
  \savebox{\mybox}{\hbox{\kern3pt$\scriptstyle#1$\kern3pt}}%
  \savebox{\mysim}{\hbox{$\sim$}}%
  \mathbin{\overset{#1}{\kern\z@\resizebox{\wd\mybox}{\ht\mysim}{$\sim$}}}%
}
\makeatother

\usepackage{changepage}
\definecolor{shadecolor1}{rgb}{0.40,0.83,0.70}
\definecolor{shadecolor2}{rgb}{0.84,0.83,0.63}

 {\endMakeFramed}

\newenvironment{shaded2}{%
  \MakeFramed {\FrameRestore}}%
 {\endMakeFramed}

\newcommand{\xval}[2]{
\vspace{.5cm}
\begin{shaded2}
\begin{quote}{\em
\footnotesize{
#1
}}\end{quote}
\footnotesize{
#2
}\end{shaded2}
\vspace{0.4cm}
}

\newenvironment{rema}
  {\begin{shaded2}}
  {\end{shaded2}}

\title{Markov Chain Monte Carlo Methods, a survey with some frequent misunderstandings}
\author{{\sc Christian P.~Robert$^{1,2}$ and
Wu Changye$^1$\footnote{This chapter is partly based on material found in the PhD
thesis of the second author, which he successfully defended in 2018 at Université Paris Dauphine,
under the supervision of the first author. Another related book chapter by the same authors 
is scheduled to appear in Mengersen, Pudlo and Robert (2020). The first author
is grateful to Antonietta Mira for her comments. Ce travail a bénéficié d'une
aide de l’Etat gérée par l'Agence Nationale de la Recherche au titre du
programme d’Investissements d’Avenir portant la référence ANR-19-P3IA-0001.}}\\
{\em $^1$Université Paris Dauphine PSL and $^2$University of Warwick}}
\date{}

\begin{document}
\maketitle

\begin{abstract}
In this chapter, we review some of the most standard tools used in Bayesian
computation, along with vignettes on standard misunderstandings of these approaches
taken from Q \&~A's on the forum Cross-validated answered by the first author.
\end{abstract}

\noindent
{\bf Acronyms:} EM, MCMC, QMC, ABC, SMC, PDMP, HMC, NUTS, PMC.

\section{Introduction}

When analysing a complex probability distribution or facing an unsolvable integration
problem, as in most of Bayesian inference, Monte Carlo methods on a large
variety of solutions, mostly based on the ability to simulate a sequence of random
variables and subsequently call for the law of large numbers. Techniques based on
the simulation of Markov chains are a special case of these methods, in which the
current simulation value (and its probability) are used to switch to a different simulation
value (hence the Markovian nature of such techniques). While the working
principle of MCMC methods was proposed almost as early as the original Monte
Carlo algorithms, the variety and efficiency of these methods has grown significantly
since \cite{gelfand:smith:1990} (re)introduced them to the statistical community
and in particular to its Bayesian component \citep{berger:1985}.

Given a likelihood function defined as a function of the parameter associated
with the probability mass function or density function of the observations $(xi^\text{obs})$,
$L(\theta|x^\text{obs})$, a Bayesian approach means relying on a so-called prior distribution on
the parameters, from which the resulting posterior distribution defined by
\begin{equation}\label{eq:posh}
\pi(\theta|x^\text{obs}) = \frac{{L}(\theta|x^\text{obs})\pi(\theta)}{\int_{\Theta}{L}(\theta|x^\text{obs})\pi(\theta')\text{d}\theta'}
\end{equation}
is derived. The denominator is sometimes called the marginal likelihood and is denoted
by $m_{\pi}(x^\text{obs})$. While most Bayesian procedures are by nature uniquely defined,
the practice of this theory exposes various computational problems.

%Insert #1
\xval{``Why is it necessary to sample from the posterior distribution if we
already know the posterior distribution?" [cross-validated:307882]}{
When one states that we ``know the posterior distribution", the meaning of
knowledge is unclear. ``Knowing" a function of $\theta$ to be
proportional to the posterior density, namely
\begin{equation}\label{eq:prod}
\pi(\theta)f(x^\text{obs}|\theta)
\end{equation}
as for instance the completely artificial following target
$$\pi(\theta|x)\propto\exp\{-||\theta-x||^2-||\theta+x||^4-||\theta-2x||^6\},\
\ x,\theta\in\mathbb{R}^{18},$$
does not mean a quick resolution for approximating the following entities
\begin{itemize}
 \item the posterior expectation of a function of $\theta$, e.g.,
$\mathbb{E}[\mathfrak{h}(\theta)|x]$, posterior mean that operates as a
Bayesian estimator under standard losses; \item the optimal decision under an
arbitrary utility function, decision that minimises the expected posterior
loss;
 \item a 90\%~or 95\%~range of uncertainty on the parameter(s), a sub-vector of
the parameter(s), or a function of the parameter(s), aka HPD
region $\{h=\mathfrak{h}(\theta);\ \pi^\mathfrak{h}(h)\ge \underline{h}\}$ 
where $\pi^\mathfrak{h}(\cdot)$ denotes the marginal posterior distribution of
$\mathfrak{h}$;
\end{itemize}
The above quantities are only examples of the infinity of usages made of a
posterior distribution. In all cases but the most simple ones, the answers are
mathematically derived from the posterior but cannot be found without
analytical or numerical steps, like Monte Carlo and Markov chain Monte Carlo
(MCMC) methods.}

The existing solutions to this computing challenge are roughly divisible into
deterministic and stochastic approaches. The former include Laplace's approximation,
expectation propagation \citep{gelman:vehtari:etal:2014} and Bayesian variational methods
\cite{jaakkola:jordan:2000}. The resulting approximation error then is usually
unknown and cannot be corrected from additional calculations. The alternative
of Monte Carlo methods leads to approximations that converge when the
computational effort becomes infinite. We will focus on the latter.

%Insert #2
\xval{``Why is variational Bayesian mixture model an alternative to MCMC? What
are the similarities?" [cross-validated:386093]}{Variational Bayes inference
is a weak form of empirical Bayesian inference \citep{berger:1985}, in the
sense that it estimates some parameters of the prior from the data for a
simplified version of the true posterior, most often of a conjugate form. The
variational Bayes approach to a Bayesian latent variable model
\citep{jaakkola:jordan:2000} is producing a pseudo-posterior distribution on
the parameters of the model, including the latent variables $\mathbf{Z}$, by
imposing a certain dependence structure (or graphical model) and estimating its
hyperparameters of this model by a maximising algorithm akin to the EM
algorithm \citep{dempster:laird:rubin:1977}.

There is thus no clear direct connection with MCMC, since the variational Bayes
posterior is made of standard distributions, thus does not require simulation,
but has hyperparameters that must be derived by an
optimisation program, hence the call to an EM-like algorithm.} 

\section{Monte Carlo methods} 
Monte Carlo approximations \citep{robert:casella:2004} are based on the Law of
Large Numbers (LLN) in the sense that an integral like
\begin{equation*}
I_h := \mathbb{E}_P(h(X))
\end{equation*}
is the limiting value of an empirical average
\begin{equation*}
\frac{1}{N} \sum_{i=1}^N h(x_i) \xrightarrow[N\rightarrow \infty]{P} I_h
\end{equation*}
when $x_1, x_2, \cdots, $ are i.i.d.~random variables with probability
distribution $P$.  In practice, the sample $x_1, x_2, \cdots, $ is produced by
a pseudo-random generator \citep{rubinstein81}

%Insert #2.3
\xval{``How can you draw samples from the posterior distribution without
first knowing the properties of said distribution? [cross-validated:307882]}
{In Bayesian settings, Monte Carlo methods are based on the assumption that the product
\eqref{eq:prod} can be numerically computed (hence
is known) for a given $(\theta,x^\text{obs})$, where $x^\text{obs}$ denotes the observation,
$\pi(\cdot)$ the prior, and $f(x^\text{obs}|\theta)$ the likelihood. This does
not imply an in-depth knowledge about this function of $\theta$. Still, from a
mathematical perspective the posterior density is completely and entirely determined by Bayes' formula,
hence derived from the product \eqref{eq:prod}.
%\begin{equation}\label{rouge}
%\pi(\theta|x^\text{obs})=\dfrac{\pi(\theta)f(x^\text{obs}|\theta)}
%{\int_ \Theta \pi(\theta)f(x^\text{obs}|\theta)\,\text{d}\theta}
%\end{equation}
Thus, it is not particularly surprising that simulation methods can be found
using solely the input of the product \eqref{eq:prod}.
The most amazing feature of Monte Carlo methods is that
some methods like Markov chain Monte Carlo (MCMC) algorithms do not
formally require anything further than this computation of the product, when
compared with accept-reject algorithms for instance, which call for an upper
bound. A related software like Stan \citep{carpenter:etal:2017} operates on this input and still
delivers high end performances with tools like NUTS \citep{hoffman2014no} and HMC, including numerical differentiation.

A side comment is that the normalising constant of the posterior \eqref{eq:posh}
%\begin{equation}\label{blue}
%\mathfrak{Z}=\int_ \Theta\pi(\theta)f(x^\text{obs}|\theta)\,\text{d}\theta
%\end{equation}
is not particularly useful for conducting Bayesian inference in that, were one to ``know" 
its exact numerical value in addition to the product \eqref{eq:prod},
$\mathfrak{Z}=3.17232\,10^{-23}$ say, one would not have made any progress
towards finding Bayes estimates or credible regions. (The only exception when
this constant matters is in conducting Bayesian model comparison.)}

\begin{rema}
\footnotesize
\begin{quote}{\em ``If we do not know the normalising constant for a posterior
distribution, why does it imply we can only sample dependent draws?"
[cross-validated:182525]}\end{quote}
This is mostly unrelated: missing normalising constant and dependence
have no logical connection. That is to say, one may have a completely
defined density and yet be unable to produce i.i.d.~samples, or one may have
a density with a missing constant and nonetheless be able to produce i.i.d.~samples.

If one knows a density $f(\cdot)$ up to a normalising constant, $f(x)\propto
p(x)$, there are instances when one can draw independent samples, using for
instance [accept-reject algorithms][1]: if one manages to find another density
$g$ such that
\begin{enumerate}
 \item one can simulate from $g$
 \item there exists a known constant $M$ such that$$p(x)\le Mg(x)$$
\end{enumerate}
then the algorithm
\begin{verbatim}
    Repeat
      simulate y~g(y)
      simulate u~U(0,1)
    until u<p(y)/Mg(y)
\end{verbatim}
\noindent produces i.i.d.~simulations from $f$, even though one only knows $p$.

For instance, if one wants to generate a Beta $\mathcal Be(a+1,b+1)$
distribution from scratch (with $a,b\ge 1$), the density up to a normalising
constant is
$$p(x)=x^a(1-x)^b\mathbb{I}_{(0,1)}(x)$$
which is bounded by $1$. Thus, we can use $M=1$ and $g(x)=1$, the density of
the uniform distribution in an accept-reject algorithm
%\begin{verbatim}
    %a=2.3;b=3.4
    %N=1e6
    %y=runif(N);u=runif(N)
    %x=y[u<y^a*(1-y)^b]
%\end{verbatim}
%
%\noindent 
that produces a sample (with random size) that is i.i.d.~from the Beta
$\mathcal Be(3.3,4.4)$ distribution. In practice, finding such a $g$ may prove
a formidable task and an easier approach is to produce simulations (asymptotically)
from $f$ by MCMC algorithms.
\end{rema}

When direct simulation from $P$, for instance a posterior distribution, is
impossible, alternative stochastic solutions must be sought. A wide
collection of such methods goes under the name of importance sampling, relying
on a convenient if somewhat arbitrary auxiliary distribution.

\begin{figure}[ht]
\centerline{\includegraphics[width=.5\textwidth]{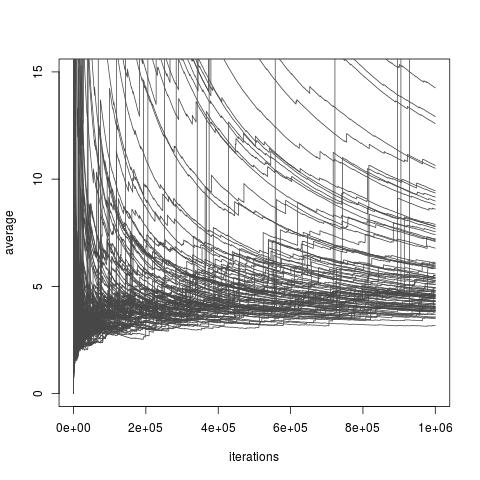}}
\caption{\small Importance sampling with importance distribution an exponential $\mathcal E(1)$
distribution, target distribution a $\mathcal E(1/10)$ distribution, and function of
interest $h(x)=x$. The true value of the expectation is equal to $10$.
The hundred curves produced on this graph correspond to repeated simulations experiments, 
with each curve describing the evolution of the empirical average of the $h(X_i)$'s with
the number of iterations. In this particular case, the importance sampling
estimators have infinite variance.\label{fig:IS}}
\end{figure}

%Insert #2.2
\xval{"What is importance sampling? [cross-validated:254114]}
{The intuition behind importance sampling is that a well-defined integral, like
$$\mathfrak{I}=\int_\mathfrak{X} h(x)\,\text{d}x$$
can be expressed as an expectation for a wide range of probability distributions with density $f$:
$$\mathfrak{I}=\mathbb{E}_f[H(X)]=\int_\mathfrak{X} H(x)f(x)\,\text{d}x$$
where $H$ is determined by $h$ and $f$. (Note that $H(\cdot)$ is usually different from
$h(\cdot)$.) The choice
$$H(x)={h(x)}\big/{f(x)}$$
leads to the equalities $H(x)f(x)=h(x)$ and
$\mathfrak{I}=\mathbb{E}_f[H(X)]$$-$under some restrictions on the support of
$f$, meaning $f(x)>0$ when $h(x)\ne 0$--. Hence, 
there is no unicity in the representation of an integral as an
expectation, but on the opposite an infinite array of such representations,
some of which are better than others once a criterion to compare them is
adopted. For instance, it may mean choosing $f$ towards reducing
the variance of the estimator.

Once this elementary property is understood, the implementation means
simulating--via a pseudo-random generator--an i.i.d.~sample $(x_1,\ldots,x_n)$
distributed from $f$ and using the average of the $H(x_i)$ as an unbiased approximation,
$\hat{\mathfrak{I}}$.  Depending on the choice of the distribution $f$, this estimator
$\hat{\mathfrak{I}}$ may or may not have a finite variance. However, there
always exist choices of $f$ that allow for a finite variance and even for an
arbitrarily small variance (albeit those choices may be unavailable in
practice). And there also exist choices of $f$ that make the importance
sampling estimator $\hat{\mathfrak{I}}$ a very poor approximation of
${\mathfrak{I}}$. This includes all the choices where the variance gets
infinite, even though \cite{chatterjee:diaconis:2018} compare importance samplers with
infinite variance. Figure \ref{fig:IS} is taken from the first author's blog discussion of the 
paper and illustrates the poor convergence of infinite variance estimators.}

A decisive appeal in using importance sampling is that the weight function w can
be known up to a multiplicative constant, which most often occurs when sampling
from a given posterior in Bayesian inference. Indeed, the multiplicative constant
can be estimated by $\frac{1}{N}\sum_{i=1}^Nw(X_i)$ and it is straightforward to deduce that the
normalised (if biased) estimator 
\begin{equation*}
{\sum_{i=1}^Nh(X_i)w(X_i)}\Big/{\sum_{i=1}^Nw(X_i)} 
\end{equation*} 
consistently approximates the integral of interest.

The importance distribution Q selected for the associated approximation significantly
impacts the quality of the method. The sequence of pseudo-random variables
that stands at the core of the method remains at this stage i.i.d. but the next section
describes a new class of sampling algorithms, based on Markov chains, which
produce correlated samples to approximate the target distribution or the integrals
of interest.

The term ``sampling" is somewhat confusing in that it does not intend to
provide samples from a given distribution.  

\begin{figure}
\centering
\begin{minipage}{.4\textwidth}
\centerline{\includegraphics[width=.95\textwidth]{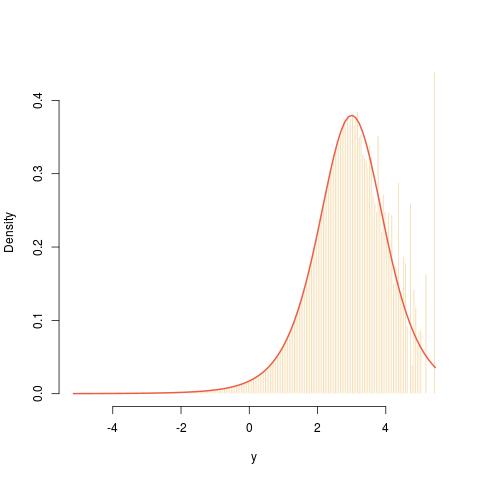}}
\caption{\label{fig:ISnot} 
\small Failed simulation of a Student's $t_5$ distribution with mean $3$ when simulating $10^7$ realisations from a standard Normal importance distribution (with thinner tails).}
\end{minipage}
\begin{minipage}{.4\textwidth}
\centerline{\includegraphics[width=.95\textwidth]{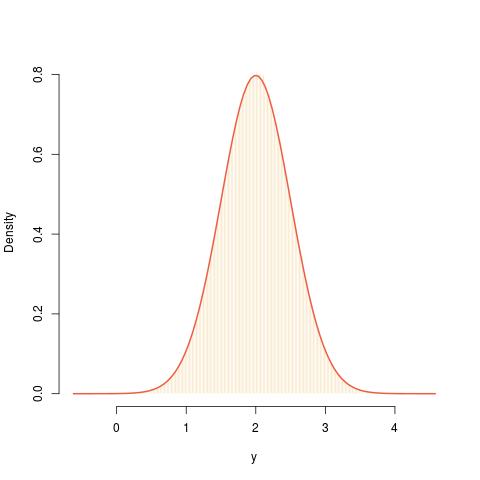}}
\caption{\label{fig:ISyes} 
\small Recovery of a Normal $\mathcal N(2,1/\sqrt{2})$ distribution when simulating $10^7$ realisations from a standard Normal importance distribution (with fatter tails).}
\end{minipage}
\end{figure}

\xval{Can importance sampling be used as an actual sampling mechanism?"[cross-validated:436453]}
{The difficulty is that the resulting (re)sample is not marginally distributed from $p$. While 
$$\mathbb E_q[h(Y) p(Y)/q(Y)]=\mathbb E_p[h(Y)]$$
for any integrable function $h(\cdot)$, weighting and resampling an i.i.d.~sample
$(Y_1,\ldots,Y_n)$ from $q$ does not produce a sample distributed from $p$, even
marginally. The reason for the discrepancy is that the weighting-resampling
step implies dividing the $p(Y_i)/q(Y_i)$ by the random sum of the weights,
i.e., the index $i$ is selected with probability
$$p(Y_i)/q(Y_i)\Big/\sum_j p(Y_j)/q(Y_j)$$
which modifies the marginal distribution of the resampled rv's, especially when the sum has an infinite variance.

Figure \ref{fig:ISnot} provides an illustration when $p$ is the density of a Student's $t_5$ distribution with mean 3 and $q$ is the density of a standard Normal distribution. The original Normal sample fails to cover the rhs of the tail of the Student's $t$ and hence that the weighted-resampled sample cannot recover with a manageable number of simulations. Obviously, as shown in Figure \ref{fig:ISyes}, when the target $q$ has fatter tails than $p$, the method converges reasonably fast.}

\begin{figure}
\centerline{\includegraphics[width=.5\textwidth]{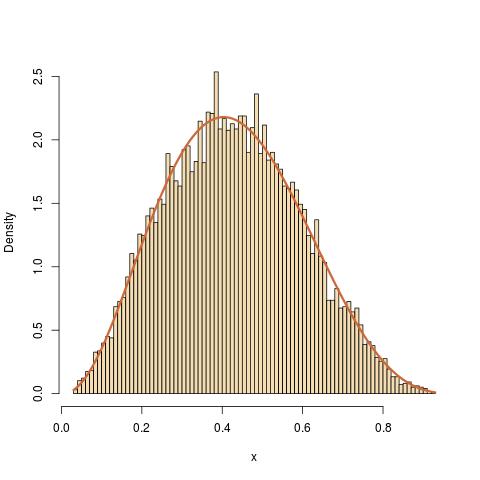}}
\caption{\label{fig:Betacpt}
\small Histogram of $9,781$ simulations of a $\mathcal
Be(3.3,4.4)$ distribution with the target density in superposition. The sample size $9,781$
is a random realisation, due to the underlying resampling mechanism.}
\end{figure}

\xval{``What is the difference between Metropolis Hastings, Gibbs, Importance, and Rejection sampling?" [cross-validated:185921]}
{These methods all produce samples from a given distribution, with
density $f$ say, either to get an idea about this distribution, or to solve an
integration or optimisation problem related with $f$. Instances include finding the
value of $$\int_{\mathcal{X}} h(x) f(x)\text{d}x\qquad h(\mathcal{X})\subset
\mathbb{R}$$ or the mode of the distribution of $h(X)$ when $X\sim f(x)$ or a
quantile of this distribution.

Here are a few generic points that do not cover the complexity of the issue:
\begin{enumerate}
\item {\em Accept-reject methods} are intended to provide an i.i.d. sample from
$f$i, as explained above. The {\em pros} are that there is no approximation in the
method: the outcome is truly an i.i.d. sample from $f$. The {\em cons} are many:
(i) designing the algorithm by finding an envelope of $f$ that can be generated
may be very costly in human time; (ii) the algorithm may be inefficient in
computing time, i.e., requires many uniforms to produce a single $x$; (iii)
those performances are decreasing with the dimension of $X$. In short, such
methods cannot be used for simulating one or a few simulations from $f$ unless
they are already available in a computer language like R.
\item {\em Markov chain Monte Carlo (MCMC) methods} are extensions of i.i.d.
simulations methods when i.i.d. simulation is too costly. They produce a
sequence of simulations $(x_t)_t$ which limiting distribution is the
distribution $f$. The {\em pros} are that (i) less information about $f$ is needed
to implement the method; (ii) $f$ may be only known up to a normalising
constant or even as an integral
$$f(x)\propto\int_{\mathcal{Z}} \tilde{f}(x,z)\text{d}z$$ 
and still be associated with an MCMC method; (iii)
there exist generic MCMC algorithms to produce simulations $(x_t)_t$ that
require very little calibration; (iv) dimension is less of an issue as large
dimension targets can be broken into conditionals of smaller dimension (as in
Gibbs sampling). The {\em cons} are that (i) the simulations $(x_t)_t$ are
correlated, hence less informative than i.i.d. simulations; (ii) the validation
of the method is only asymptotic, hence there is an approximation in
considering $x_t$ for a fixed $t$ as a realisation of $f$; (iii) convergence to
$f$ (in $t$) may be so slow that for all practical purposes the algorithm does
not converge; (iv) the universal validation of the method means there is an
infinite number of potential implementations, with an equally infinite range of
efficiencies. 
\item {\em Importance sampling methods} are originally designed for integral
approximations, namely generating from the wrong target $g(x)$ and compensating
by an importance weight $f(x)/g(x)$. The resulting sample is thus weighted,
which makes the comparison with the above awkward. Importance sampling
can be turned into importance sampling resampling by using an additional
resampling step based on the weights, still failing to produce an exact simulation
from the target as discussed above. The {\em pros} of importance sampling
are that (i) generation from an importance target $g$ can be cheap
and recycled for different targets $f$; (ii) the ``right" choice of $g$ can lead
to huge improvements compared with regular or MCMC sampling; (iii) importance
sampling is more amenable to numerical integration improvement, like for
instance quasi-Monte Carlo integration; (iv) it can be turn into adaptive
versions like population Monte Carlo and sequential Monte Carlo. The {\em cons} are
that (i) resampling induces inefficiency (which can be partly corrected by
reducing the noise as in systematic resampling or qMC); (ii) the ``wrong" choice
of $g$ can lead to huge losses in efficiency and even to infinite variance;
(iii) importance has trouble facing large dimensions and its efficiency
diminishes quickly with the dimension; (iv) the method may be as myopic as
local MCMC methods in missing important regions of the support of $f$.
\end{enumerate}

A final warning is that {\em there is no such thing as an optimal
simulation method.} Even in a specific setting like approximating an integral
$\mathcal{I}$, costs of designing and
running different methods intrude as to make a global comparison very delicate,
if at all possible, while, from a formal point of view, {\em they can never beat
the zero variance answer of returning the constant "estimate"}. For instance,
simulating from $f$ is very rarely if ever the best option. This does not mean
that methods cannot be compared, but that there always is a possibility for an
improvement, which usually comes with additional costs.}

\section{Markov chain Monte Carlo methods}

\noindent 
Markov chain Monte Carlo (MCMC) algorithms are now standard computing tools
for analysing Bayesian complex models \cite{gelfand:smith:1990} even though practitioners
may still face difficulties with its implementations. The concept behind
MCMC is quite simple in that it creates a sequence of dependent variables that
converge (in distribution) to the distribution of interest (also called target). In that
sense, MCMC algorithms are robust or universal, as opposed to the most standard
Monte Carlo methods which require direct simulations from the target distribution.

%Insert #3
\xval{``Is Markov chain based sampling the ``best'' for Monte Carlo sampling?
Are there alternative schemes available?" [cross-validated:131455]}
{There is no reason that MCMC sampling is the ``best" Monte Carlo
method! Usually, it is on the opposite {\em worse} than i.i.d.~sampling, at least
in terms of variance of the resulting Monte Carlo
estimators$$\frac{1}{T}\sum_{t=1}^T h(X_t)$$Indeed, while this average
converges to the expectation $\mathbb{E}_{\pi}[h(X)]$ when $\pi$ is the
stationary and limiting distribution of the Markov chain $(X_t)_t$, there are
at least two drawbacks in using MCMC methods:
\begin{enumerate}
\item The chain needs to ``reach stationarity", meaning that it needs to forget
about its starting value $X_0$. In other words, $t$ must be ``large enough" for
$X_t$ to be distributed from $\pi$. Sometimes ``large enough" may exceed by
several orders of magnitude the computing budget available for the experiment.
\item The values $X_t$ are correlated, leading to an asymptotic variance that involves
$$\text{var}_\pi(X)+2\sum_{t=1}^\infty\text{cov}_\pi(X_0,X_t)$$ 
which generally exceeds $\text{var}_\pi(X)$ and hence requires longer simulations 
than for an i.i.d.~sample, as well as more involved evaluation techniques.
\end{enumerate}

This being said, MCMC is very useful for handling settings where regular i.i.d.~sampling is impossible or too costly and where importance sampling is quite
difficult to calibrate, in particular because of the dimension of the random
variable to be simulated.  However, sequential Monte Carlo methods \citep{liu:chen:logvinenko:2001} 
like particle filters may be more appropriate in dynamical models, where the data
comes by bursts that need immediate attention and may even vanish (i.e., cannot
be stored) after a short while.}

From the early 1950's, MCMC methods \citep[see, e.g.][]{cappe:robert:2000b,robert:casella:2010,green:etal:2015} have been utilised to handle complex target
distributions by simulation, where the meaning of complexity depends on the target
density, the size of the associated data, the dimension of the object to be simulated,
or on the allocated budget. For instance, the density $p(x)$ is only expressed as a
multidimensional integral that is analytically intractable,
$$
p(x) = \int \omega(x,\xi)\text{d}\xi\,.
$$ 
An evaluation of this density requires the simulation of the whole vector $(x,\xi)$.

In these cases when $\xi$ has a dimension at least as large as the data, this involves
a significant increase in the dimension of the simulated object and hence deeper
computational difficulties, like handling the new target $\omega(x,\xi)$. An MCMC
algorithm provides an alternative solution to this computational issues through a
simulated Markov chain evolving in the augmented space without requiring further
information on the density $p$. 

\begin{figure}
\centering
\begin{minipage}{.45\textwidth}
\centerline{\includegraphics[width=.95\textwidth]{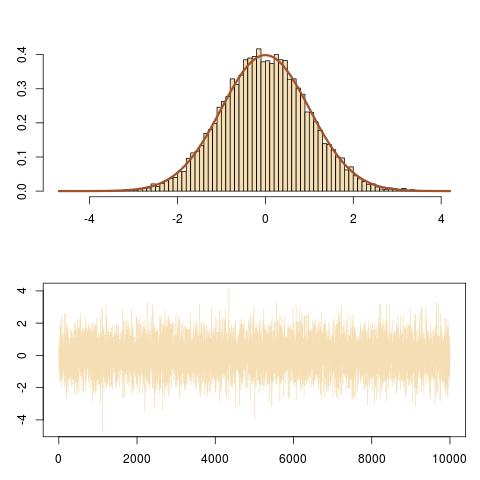}}
\caption{\label{fig:NMH}
{\em top:} Histogram of $10^4$ iterations of a slice sampler with a Normal $\mathcal N(0,1)$ target; 
{\em bottom:} sequence $(X_i)$}
\end{minipage}
\begin{minipage}{.45\textwidth}
\centerline{\includegraphics[width=.95\textwidth]{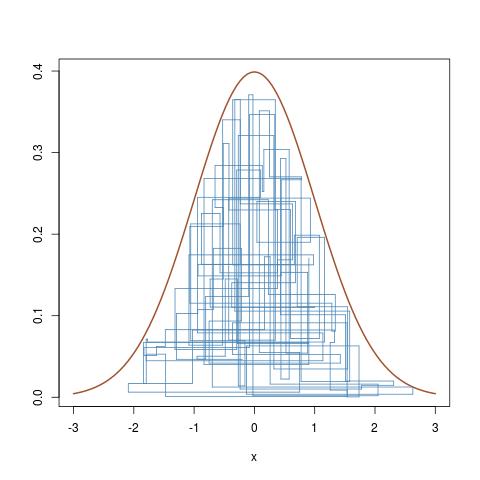}}
\caption{\label{fig:zooNHM}
100 last moves of the above slice sampler.}
\end{minipage}
\end{figure}

%Insert #3.1
\begin{rema}
\footnotesize
\begin{quote}{\em
What is the connection between Markov chain and Markov chain Monte Carlo?"[cross-validated:169518]}
\end{quote}
The connection between both concepts is that Markov chain Monte Carlo (MCMC) methods] rely on Markov chain theory to produce
simulations and Monte Carlo approximations from a complex target distribution $\pi$.

In practice, these simulation methods output a sequence $X_1,\ldots,X_N$ that
is a Markov chain, i.e., such that the distribution of $X_i$ given the whole
past $\{X_{i-1},\ldots,X_1\}$ only depends on $X_{i-1}$. In other words,
$$X_i=f(X_{i-1},\epsilon_i)$$ 
where $f$ is a function specified by the
algorithm and the target distribution $\pi$ and the $\epsilon_i$'s are i.i.d.. The
(ergodic) theory guarantees that $X_i$ converges (in distribution) to $\pi$ as
$i$ gets to $\infty$.

The easiest example of an MCMC algorithm is the slice sampler: at iteration $i$ of this algorithm, do
\begin{enumerate}
\item simulate $\epsilon^1_i\sim\mathcal{U}(0,1)$
\item simulate $X_{i}\sim\mathcal{U}(\{x;\pi(x)\ge\epsilon^1_i\pi(X_{i-1})\})$ (which
amounts to generating a second independent $\epsilon^2_i$)
\end{enumerate}
For instance, if the target is a Normal $\mathcal{N}(0,1)$ distribution\footnote{For
which one obviously does not need MCMC in practice: this is a toy example.}
the above translates as
\begin{enumerate}
\item simulate $\epsilon^1_i\sim\mathcal{U}(0,1)$
\item simulate $X_{i}\sim\mathcal{U}(\{x;x^2\le-2\log(\sqrt{2\pi}\epsilon^1_i\})$,
i.e.,$$X_i=\pm\epsilon_i^2\{-2\log(\sqrt{2\pi}\epsilon^1_i)\varphi(X_{i-1})\}^{1/2}$$
with $\epsilon_i^2\sim\mathcal{U}(0,1)$
\end{enumerate}
%or in R
%\begin{verbatim}
    %T=1e4
    %x=y=runif(T) #random initial value
    %for (t in 2:T){
      %epsilon=runif(2)#uniform white noise
      %y[t]=epsilon[1]*dnorm(x[t-1])#vertical move
      %##Markov move from x[t-1] to x[t]
      %x[t]=sample(c(-1,1),1)*epsilon[2]*sqrt(-2*
            %log(sqrt(2*pi)*y[t]))}
%\end{verbatim}
Figure \ref{fig:NMH} is a representation of the output, showing the right fit
to the $\mathcal{N}(0,1)$ target and the evolution of the Markov chain $(X_i)$.
And Figure \ref{fig:zooNHM} zooms on the evolution of the Markov chain
$(X_i,\epsilon^1_i\pi(X_i))$ over the last 100 iterations,
%obtained by
%\begin{verbatim}
    %curve(dnorm,-3,3,lwd=2,col="sienna",ylab="")
    %for (t in (T-100):T){
    %lines(rep(x[t-1],2),c(y[t-1],y[t]),col="steelblue");
    %lines(x[(t-1):t],rep(y[t],2),col="steelblue")}
%\end{verbatim}
which follows vertical and horizontal moves of the Markov chain under the target density curve.
\end{rema}

The validation of the method \citep[e.g.,][]{robert:casella:2004} proceeds by
establishing that the resulting Markov chain is ergodic
\citep[e.g.,][]{meyn:tweedie:1993}, meaning that it converges 
to the distribution corresponding to $\pi$, making the starting
value of the chain irrelevant. Akin to basic Monte Carlo methods, MCMC samples
(usually) enjoy standard limit theorems.

\subsection{Metropolis-Hastings algorithms}
\noindent The Metropolis--Hastings\footnote{in reference to N.~Metropolis, with whom the
algorithm originated \citep{metropolis:1953} and K.~Hastings, for his generalisation \citep{hastings1970monte}
} algorithm is the ``Swiss knife" of MCMC methods in that
it offers a form of universal solution to the construction of an appropriate Markov
chain. The algorithm requires a proposal distribution, with density $q(x'|x)$ and
proceeds one step at a time based on simulations proposed from this distribution
and accepted or rejected by a Metropolis--Hastings ratio, as described in Algorithm
\ref{algo:MHzero}.
%$$X_{n+1} = \begin{cases}
%X^\prime \sim q(X^\prime |X_n) &\text{with probability }
%\left\{\dfrac{p(X^\prime)}{p(X_n)}\times\dfrac{q(X_n|X^\prime)}
%{q(X^\prime|X_n)}\right\}\wedge 1,\\
%X_n &\text{otherwise}.\\
%\end{cases}$$
\begin{algorithm}[H]
\caption{\small Metropolis-Hastings algorithm}
\footnotesize
\begin{algorithmic}\label{algo:MHzero}
\STATE{\textbf{Input}: starting point $X_0$, proposal distribution $q$ and number of iterations $N$.}
\FOR{$n = 1,2,\cdots, N$}
\STATE{Sample $X' \sim q(\cdot|X_{n-1})$}
\STATE{Compute the acceptance probability $\alpha(X_{n-1}, X')$, where}
\STATE{\begin{equation*}
\alpha(X_{n-1}, X') = \min\left\{1, {p(X')q(X_{n-1}|X')}\Big/{p(X_{n-1})q(X'|X_{n-1})}\right\}
\end{equation*}}
\STATE{Sample $U\sim \mathcal{U}[0,1]$;}
\IF{$U < \alpha(X_{n-1}, X')$}
     \STATE{$X_n \rightarrow X'$ }
\ELSE
     \STATE{$X_n \rightarrow X_{n-1}$}
\ENDIF
\ENDFOR
\end{algorithmic}
\end{algorithm}
\noindent The accept--reject step in this algorithm is fundamental in that it turns $p$ into its
stationary distribution, assuming the resulting Markov kernel is irreducible, provided
the chain $(X_n)_n$ is irreducible, meaning it has a positive probability of hitting any
part of the support of $p$ on a finite number of steps. Stationary follows from the
transition satisfying the detailed balance condition, corresponding to the chain being
reversible in time, \citealp[see, e.g.,][]{robert:casella:2004}.
A special case when $q$ is symmetric, i.e., $q(x|y) = q(y|x)$ is called random walk MCMC and the
acceptance probability only involves the targeted $p$.

\xval{``What is the deeper intuition behind the symmetric proposal distribution in the Metropolis-Hastings Algorithm?" \ [cross-validated:262216]
\begin{enumerate}
\item the Normal and Uniform are symmetric probability density functions
themselves, is this notion of ``symmetry" the same as the ``symmetry" above?
\item is there an intuitive way of seeing the deeper meaning behind the symmetry formula above?
\end{enumerate}}{
Both Normal and Uniform distributions are symmetric around their mean. But the symmetry in
Metropolis-Hastings signifies that $q(x|y)=q(y|x)$ which makes the ratio cancel in the
Metropolis-Hastings acceptance probability. If one uses a Normal distribution
not centered at the previous value in the Metropolis-Hastings proposal (as e.g.
in the Langevin version), the Normal distribution remains symmetric {\em as a
distribution} but the proposal distribution is no longer symmetric and hence it
must appear in the Metropolis-Hastings acceptance probability.

There is no particular depth in this special symmetric case, it simply makes
life easier by avoiding the ratio of the proposals. It may save time or it may
avoid computing complex or intractable densities. Note also that the symmetry
depends on the parameterisation of the model: if one changes the
parameterisation, a Jacobian appears and kills the symmetry.}

\begin{figure}
\centerline{\includegraphics[width=.5\textwidth]{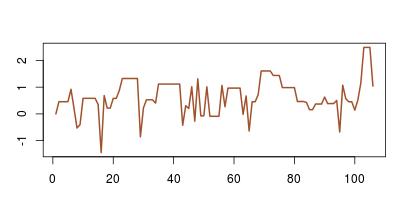}}
\caption{\label{fig:sYsN} 
\small Independent Metropolis sequence with a proposal $f_V$ equal to the
density of a $\mathcal N(0,1)$ distribution and a target $f_Y$ being the
density of a $\mathcal N(1,1)$ distribution.}
\end{figure}

\xval{The independent Metropolis Algorithm using the proposal $X'\sim f_V(x)$
should have $\alpha(X_0,X_0') = 1$ and hence the chain always equal to $X_0'$.
[cross-validated:396704]}{
The confusion stems from a misunderstanding of the notation $X' \sim f_V$
which means both (a) $X'$ is a random variable with density $f_V$ and (b) $X'$ is
created by a pseudo-random generation algorithm that reproduces a generation of a random
variable with density $f_V$. Each time a generation $X_i'\sim f_V$ occurs in the
algorithm, a new realisation of a random variable
with density $f_V$ occurs, which is independent from all previous realisations,
hence different from these previous realisations. Equivalently, stating that the
$X_i'$ are all identically distributed from the same distribution $f_V$ does
not mean that their realisations all are numerically identical.

The starting point of the Metropolis-Hastings algorithm is arbitrary, either
fixed $X_0=0$ for instance or random, for instance $X_0\sim f_V$ (a notation
meaning that $X_0$ is distributed from $f_V$). This starting value {\em is
always accepted}. For $i=1$, one generates $X_1'\sim f_V$ (meaning that $X_1'$ is
distributed from $f_V$, independently and thus different from $X_0$)
$$X_1 =\begin{cases}
      X_1' & \text{if }U_1\leq \alpha_1=\min\left(\frac{f_Y(X_1')}{f_V(X_1')} \frac{f_V(X_{0})}{f_Y(X_{0})},1\right) \\
      X_{0} & \text{if }U_1 > \alpha_1
   \end{cases}$$
and $\alpha_1\ne 1$ in general. Hence sometimes $X_1$ is accepted and sometimes
not. The same applies to the following steps.  To make a toy illustration on
how the algorithm applies, take $f_V$ to be the density of a $\mathcal N(0,1)$
distribution and $f_Y$ to be the density of a $\mathcal N(1,1)$ distribution. A
sequence of i.i.d.~generations from $f_V$ is for instance (by a call to R {\sf rnorm})
$$0.45735433,-0.99178415,-1.08312586,-0.85762451,0.92186197,-0.50442298,...$$
(note that they are all different) and a sequence of generations from $\mathcal
U$ is for instance (by a call to R {\sf runif})
$$0.441328,0.987837,0.386258,0.316593,0.195910,0.2772669,...$$
(note that they are all different). Applying the algorithm with starting value
$X_0=0$ means considering
$$\frac{f_Y(X_1')}{f_V(X_1')} \frac{f_V(X_{0})}{f_Y(X_{0})}=0.9582509\big/0.6065307=1.579889>1$$
which implies that $X_1=X_1'=0.45735433$. Then
$$\frac{f_Y(X_2')}{f_V(X_2')} \frac{f_V(X_{1})}{f_Y(X_{1})}= 0.2249709 \big/ 0.9582509 = 0.2347724 < U_2=0.987837$$
which implies that $X_2=X_1$. The algorithm can be applied step by step to the
sequences provided above, which leads to
\begin{align*}
\frac{f_Y(X_3')}{f_V(X_3')} \frac{f_V(X_{2})}{f_Y(X_{2})}&=0.2053581 \big/0.9582509 = 0.2143051 < U_3 \qquad Z_3 = Z_1\\
\frac{f_Y(X_4')}{f_V(X_4')} \frac{f_V(X_{3})}{f_Y(X_{3})}&=0.2572712 \big/0.9582509 = 0.2684800 < U_4 \qquad Z_4 = Z_1\\
\frac{f_Y(X_5')}{f_V(X_5')} \frac{f_V(X_{4})}{f_Y(X_{4})}&=1.5247980 \big/0.9582509 = 1.591230 > 1 \qquad\quad  Z_5 = V_5\\
\end{align*}
producing a sequence as in Figure \ref{fig:sYsN} (notice the flat episodes in the graph, which correspond to a sequence of rejections).

As a final remark, the only potentially confusing part in the description in
Casella and Berger (1990) is the very first sentence where the random variables
$Y$ and $V$ are not needed. It could have been clearer to state ``Let $f_Y$ and
$f_V$ be two densities with common support."
}

\noindent Since the purpose of MCMC methods like the Metropolis-Hastings algorithm is to
simulate realisations from $p$, their performances are highly variable. These obviously
depends on the connection between p and q. For instance, the Metropolis-Hastings
algorithm is an i.i.d.~sampler when $q(\cdot|X_n)=p(\cdot)$,
a choice that is rarely available.
Although it may happen that the Markov chain $(X_n)$ achieves negative correlations
between successive and further terms of the series, making it em more efficient than
i.i.d.~sampling \citep{liu:won:kon95}, it is more common that there exists a positive
covariance between simulations (sometimes for all transforms, see \citealp{liu:won:kon94}).
This feature means a lesser efficiency of the algorithm which thus requires a greater
number of simulations to achieve the same accuracy as the i.i.d. approach (regardless
of the differences in computing time). In general, the MCMC algorithm may require
a large number of iterations to escape the attraction of the starting point X0 and
to converge. There is a real danger that some versions of these algorithms do not
converge within the allotted time (in practice if not in theory).

\begin{figure}
\centering
\begin{minipage}{.45\textwidth}
\centerline{\includegraphics[width=.95\textwidth]{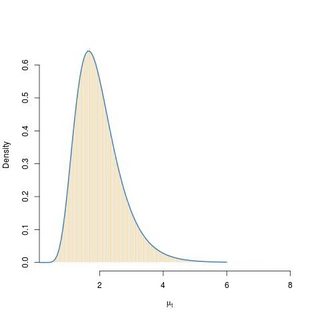}}
\caption{\label{fig:trunk}
\small
Fit of a Metropolis sample of size $10^4$ to a target when using a truncated Normal proposal.}
\end{minipage}
\begin{minipage}{.45\textwidth}
\centerline{\includegraphics[width=.95\textwidth]{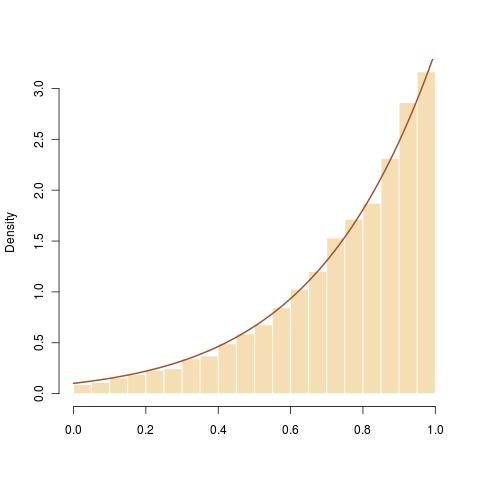}}
\caption{\label{fig:valiD}
\small
Graph of a truncated Normal density and fit by the histogram of an MCMC sample
using a Gaussian random walk.}
\end{minipage}
\end{figure}

%Insert #4
\begin{rema}
\footnotesize
\begin{quote}{\em ``What is the Metropolis-Hastings acceptance ratio for a truncated proposal?" 
[cross-validated:345291]}
\end{quote}

If a Metropolis-Hastings algorithm uses a truncated Normal as proposal, e.g., the positive Normal
$${\mathcal N}^+(\mu_{t-1},\sigma^2)$$
the associated Metropolis-Hastings acceptance ratio is
$$\dfrac{\pi(\mu')}{\pi(\mu_{t-1})}\times
\dfrac{\varphi(\{\mu_{t-1}-\mu'\}/\sigma)}{\varphi(\{\mu'-\mu_{t-1}\}/\sigma)}
\times\dfrac{\Phi(\mu_{t-1}/\sigma)}{\Phi(\mu'/\sigma)}$$
when $\mu'\sim{\cal N}^+(\mu_{t-1},\sigma^2)$ is the proposed value and $\pi$
denotes the target of the simulation (e.g., the posterior distribution). This ratio simplifies into
$$\dfrac{\pi(\mu')}{\pi(\mu_{t-1})}\times \dfrac{\Phi(\mu_{t-1}/\sigma)}{\Phi(\mu'/\sigma)}$$
hence the truncation impacts the Metropolis-Hastings acceptance ratio.

Figure \ref{fig:trunk} provides an illustration for the target density
$$\pi(\mu)\propto\exp\{-(\log \mu -1)^2\}\,\exp\{-(\log \mu -3)^4/4\}$$
when using $\sigma=.1$ as the scale in the truncated Normal. 
%obtain as
%\begin{verbatim}
    %targ=function(mu){1/exp((log(mu)-1)^2+log(mu)-3)^4/4}
    %mumc=rep(pi,1e4)
    %for (t in 2:1e4){
      %prop=mumc[t-1]+.1*qnorm(pnorm(-mumc[t-1]/.1)+
          %runif(1)*pnorm(mumc[t-1]/.1))
      %mumc[t]=prop+(runif(1)>targ(prop)*pnorm(mumc[t-1]/.1)/
           %targ(mumc[t-1])/pnorm(prop/.1))*(mumc[t-1]-prop)}
%\end{verbatim}
\end{rema}

\xval{``What to do when rejecting a proposed point in MCMC?" [cross-validated:123113]}
{
The validation of the Metropolis-Hastings algorithm relies on repeating the
current value in the Markov chain if the proposed value is rejected. One should
not consider the list of accepted* points as one's sample but instead the
Markov chain with transition
\begin{align*}
X_{t+1} &= Y_{t+1} \quad&\text{if } U_{t+1}\le \pi(Y_{t+1})/\pi(X_t)\\
&= X_t \quad&\text{otherwise}
\end{align*}
(assuming a symmetric proposal distribution). The repetition of the current
value in the event of a rejection is what makes the algorithm valid, i.e., why
$\pi$ is the stationary distribution.

It is always possible to study the distribution of the accepted and of the
rejected values, with some recycling possible by
Rao-Blackwellisation \citep{casella:robert:1996}, but this study is more advanced and far from
necessary to understand the algorithm.
}

\begin{rema}
\footnotesize
\begin{quote}
{\em How to account for impossible proposed values? [cross-validated:51808]}
\end{quote}
It is indeed a popular belief that something needs to be done to account for
restricted supports. However, there is no mathematical reason for doing so.
The Metropolis-Hastings acceptance probability
$$\rho(x_t,y_{t+1})= \text{min} (1, \pi(y_{t+1})q(x_t|y_{t+1})\big/
\pi(x_t)q(y_{t+1}|x_{t}) )$$
with $y_t\sim q(y_{t+1}|x_{t})$ can handle cases when
$y_t$ is outside the support of $\pi$ by extending this support, defining
$\pi(y)=0$ outside the original support. Hence, if $\pi(y_{t+1})=0$,
then $\rho(x_t,y_{t+1})=0$, which means the proposed value is automatically
rejected and $x_{t+1}=x_t$.

Consider the following illustration.
\begin{verbatim}
    target=function(x) (x>0)*(x<1)*dnorm(x,mean=4)
    mcmc=rep(0.5,10^5)
    for (t in 2:10^5){
    prop=mcmc[t-1]+rnorm(1,.1)
    if (runif(1)<target(prop)/target(mcmc[t-1]))
    mcmc[t]=prop
    else
    mcmc[t]=mcmc[t-1]
    }
    hist(mcmc,prob=TRUE)
    curve(dnorm(x-4)/(pnorm(-3)-pnorm(-4)),add=TRUE)
\end{verbatim}
that is targeting a truncated normal distribution using a Gaussian random walk
proposal with support the entire real line. Then the algorithm is properly
converging as shown by the fit in Figure \ref{fig:valiD}
\end{rema}

\xval{\noindent Unbiased MCMC [xianblog:25/08/2017]}{
\cite{jacob:leary:atchade:2020} propose an unbiased MCMC technique based on
coupling. Associating MCMC with unbiasedness is rather challenging since MCMC
are rarely producing simulations from the exact target, unless specific tools
like renewal can be produced in an efficient manner. 

The central idea is coupling of two (MCMC) chains, associated with the
debiasing formula used by \cite{glynn:rhee:2014}.
Having the coupled chains meet at some time with probability one implies that
the debiasing formula does not need a (random) stopping time. The coupling time
is sufficient. Furthermore, several estimators can be derived from the same
coupled Markov chain simulations, obtained by starting the averaging at a later
time than the first iteration. The average of these (unbiased) averages results
into a weighted estimate that weights more the later differences. Although
coupling is also at the basis of perfect simulation methods, the analogy
between this debiasing technique and perfect sampling is hard to fathom, since
the coupling of two chains is not a perfect sampling instant. (Something
obvious in retrospect is that the variance of the resulting
unbiased estimator is at best the variance of the original MCMC estimator.)

When discussing the implementation of coupling in Metropolis and Gibbs
settings, the authors produce a simple optimal coupling algorithm,
a form of accept-reject also found in perfect sampling. While I did not fully
understood the way two random walk Metropolis steps are coupled, in that the
normal proposals seem at odds with the boundedness constraints, coupling is
clearly working in this setting, while renewal does not. In toy examples like
the \cite{efron:morris:1973} baseball data and the \cite{gelfand:smith:1990} pump failure
data, the parameters of the algorithm can be optimised against the
variance of the averaged averages. And this approach proves highly useful in the
case of the cut distribution.}

\subsection{Gibbs sampling}

Historically, this form of MCMC algorithm is distinguished from the other types
of MCMC methods for being both justified by other arguments and used for a
specific class of models \citep{robert:casella:2004}.

\xval{``Why would one use Gibbs sampling instead of Metropolis-Hastings?" [cross-validated:244573]}
{The question does not have an answer in that a Metropolis-Hastings sampler can
be almost anything, including a Gibbs sampler.  The primary reason why Gibbs
sampling was introduced was to break the curse of
dimensionality (which impacts both rejection and importance sampling) by
producing a sequence of low dimension simulations that still converge to the
right target. Even though the dimension of the target impacts the speed of
convergence. Metropolis-Hastings samplers are designed to create a Markov chain
(like Gibbs sampling) based on a proposal (like importance and rejection
sampling) by correcting for the wrong density through an acceptance-rejection
step. But an important point is that they are not opposed: namely, Gibbs
sampling may require Metropolis-Hastings steps when facing complex if
low-dimension conditional targets, while Metropolis-Hastings proposals may be
built on approximations to (Gibbs) full conditionals. In a formal definition,
Gibbs sampling is a special case of Metropolis-Hastings algorithm with a
probability of acceptance of one. 

Usually, Gibbs sampling--understood as running a sequence of low-dimensional
conditional simulations--is favoured in settings where the decomposition into
such conditionals is easy to implement and fast to run. In settings where such
decompositions induce multimodality and hence a difficulty to move between
modes (latent variable models like mixture models come to mind), using a more
global proposal in a Metropolis-Hastings algorithm may produce a higher
efficiency. But the drawback stands with choosing the proposal distribution in
the Metropolis-Hastings algorithm.
}

\subsection{Hamiltonian Monte Carlo}
\noindent A more advanced (and still popular) form of MCMC algorithm is Hamiltonian Monte
Carlo (HMC) \citep{duane:etal:1987,neal:1996,neal:2011}. While a special case of continuous
time samplers, it can be implemented in discrete time and is actually behind
the successful Stan package \citep{carpenter:etal:2017}. The construction of the process
relies on an auxiliary variable $v$ that augments the target into
$$
\rho(x,v)
=p(x)\varphi(v|x) \propto \exp\{ -H(x,v) \}\,,
$$
where $\varphi(v|x)$ is the conditional density of v given x. This density obviously
enjoys $p(v)$ as its marginal and while it could be anything, the so-called momentum
$v$ is usually chosen of the same dimension as $v$, with $\varphi(v|x)$ often taken as a Normal
density. The associated {\em Hamiltonian equations}
$$
\dfrac{\text{d}x_t}{\text{d}t}=\dfrac{\partial H}{\partial v}(x_t,v_t)\qquad
\dfrac{\text{d}v_t}{\text{d}t}=-\dfrac{\partial H}{\partial x}(x_t,v_t)\,,
$$
which keeps the {\em Hamiltonian} target $H(\cdot)$ constant over time, as
$$
\dfrac{\text{d}H(x_t,v_t)}{\text{d}t}=\dfrac{\partial H}{\partial v}(x_t,v_t)\,\dfrac{\text{d}v_t}{\text{d}t}+\dfrac{\partial H}{\partial x}(x_t,v_t)\,\dfrac{\text{d}x_t}{\text{d}t}=0\,.
$$
Since there is no randomness in the above process, 
the HMC algorithm is completed with random changes of the momentum
according to the correct conditional distribution, $v_t\sim\varphi(v|x_t)$, at times driven by a
Poisson process $\{\tau_n\}_n$.

As noted above, the choice of the conditional density $\varphi(v|x_t)$ often is a Gaussian
density with either a constant covariance matrix $M$ calibrated from the target
covariance or as a local curvature depending on $x$ in the version of \cite{girolami:2011}
called Riemannian Hamiltonian Monte Carlo. See, e.g.,
\cite{livingstone2017kinetic} for an analysis of the impact of different types
of kinetic energy on Hamiltonian Monte Carlo performances.

When the fixed covariance matrix is equal to $M$, the Hamilton equations write as
$$
\dfrac{\text{d}x_t}{\text{d}t}=
M^{-1}v_t\qquad \dfrac{\text{d}v_t}{\text{d}t}=\nabla \log{p}(x_t)\,,
$$
where the last term is the score function. The velocity of the HMC process is thus
connected to the gradient of the log-target.

In practice, implementing this rather simple remark proves rather formidable
in that there is no direct approach for methodology for simulating this continuous
time process, since the above equations cannot are intractable. A natural resolution
associates a numerical solver like Euler's method, usually unstable, with a numerical
solver naturally suited to these equations.

This general method is called a symplectic integrator \citep{betancourt:2017} with
implementation in the constant covariance case resorting to time-discretisation leapfrog
steps
\begin{align*}
v_{t+\epsilon/2} &= v_t+\epsilon \nabla \log{p}(x_t)/2,\\
x_{t+\epsilon} &= x_t+\epsilon M^{-1} v_{t+\epsilon/2},\\
v_{t+\epsilon} &= v_{t+\epsilon/2}+\epsilon \nabla \log{p}(x_{t+\epsilon})/2,
\end{align*}
which symmetrises the two-step move, with $\epsilon$ standing for the
time-discretisation step. The proposed value of $v_0$ is generated from the
true Gaussian target. The correction to the discretisation approximation
involves a Metropolis--Hastings step over the pair $(x_{t+\epsilon} v_{t+\epsilon})$
which reintroduces some reversibility into the picture.

Time-discretising the Hamiltonian dynamics in the leapfrog integrator involves two quantities, 
$\epsilon$ and $T$ the trajectory length. One empirically sound calibration of these
parameters is found in the ``no-U-turn sampler" of \cite{hoffman2014no}
which selects the value of $N$ by primal-dual averaging and produces the trajectory
length $T$ as the length of the chain is takes for the path to fold back.

%\begin{minipage}{.45\textwidth}
\begin{algorithm}[h]
\caption{\small Leapfrog($x_0, v_0, \epsilon, L$)}
\footnotesize
\begin{algorithmic}
\STATE{\textbf{Input}: {starting position $x_0$, starting momentum $v_0$, step-size $\epsilon$, steps $L$}}
\FOR{$\ell = 0,1, \cdots, L-1$}
\STATE{$ v_{\ell+1/2} = v_{\ell}+\epsilon \nabla \log{p}(x_{\ell})$}
\STATE{$x_{\ell+1} = x_{\ell}+\epsilon M^{-1} v_{\ell+1/2}$}
\STATE{$v_{\ell+1} = v_{\ell+1/2}+\epsilon \nabla \log{p}(x_{\ell+1})$}
\ENDFOR
\STATE{\textbf{Output}: {$(x_{L}, v_{L})$}}
\end{algorithmic}
\end{algorithm}      
%\end{minipage}
%\begin{minipage}{.45\textwidth}
\begin{algorithm}[h]
\caption{\small Hamiltonian Monte Carlo algorithm}
\footnotesize
\begin{algorithmic}
\STATE{\textbf{Input}: {step-size $\epsilon$, steps of leapfrog integrator $L$, starting position $x_{0}$}, desired number of iterations $N$.}
\FOR{$n = 1, \cdots, N$}
\STATE Sample $v_{n-1} \sim \varphi(v)$;
\STATE Compute $(x^*, v^*) \leftarrow \text{Leapfrog}(x_{n-1}, v_{n-1}, \epsilon, L)$;
\STATE Compute the acceptance ratio $\alpha$, where
\STATE{\begin{equation*}
\alpha = \min\left\{1, {\exp(-H(x^*, -v^*))}\big/{\exp(-H(x_{n-1}, v_{n-1}))}\right\};
\end{equation*}}
\STATE Sample $u \sim \mathcal{U}[0,1]$;
\IF{$u <\alpha$} 
\STATE{$x_n \leftarrow x^*$}
\ELSE
\STATE{$x_n \leftarrow x_{n-1}$}
\ENDIF
\ENDFOR
\end{algorithmic}
\end{algorithm}      
%\end{minipage}
\noindent\normalfont
In practice, it is important to note that discretising Hamiltonian dynamics
introduces two free parameters, the step size $\epsilon$ and the trajectory
length $T$, both to be calibrated. As an empirically successful and popular
variant of HMC, the ``no-U-turn sampler'' (NUTS) of \cite{hoffman2014no} adapts
the value of $\epsilon$ based on primal-dual averaging. It also eliminates the
need to choose the trajectory length $T$ via a recursive algorithm that builds
a set of candidate proposals for a number of forward and backward leapfrog
steps and stops automatically when the simulated path retraces. 

\xval{\noindent Unbiased HMC [xianblog:25/09/2017]}{
\cite{hung:jacob:2019} propose to achieve unbiased Hamiltonian Monte Carlo by coupling, following 
\cite{jacob:leary:atchade:2020} discussed earlier. The coupling within the HMC amounts to running two HMC chains with common random numbers, plus subtleties.

    \begin{quote}\begin{em}``As with any other MCMC method, HMC estimators are justified in the limit of the number of iterations. Algorithms which rely on such asymptotics face the risk of becoming obsolete if computational power keeps increasing through the number of available processors and not through clock speed."\end{em}
Heng and Jacob (2019)
\end{quote}

The main difficulty here is to have both chains meet (exactly) with large
probability, since coupled HMC can only bring these chain close to one another.
The trick stands in using both coupled HMC and coupled Hastings-Metropolis
kernels, since the coupled MH kernel allows for exact meetings when the chains
are already close, after which they remain forever identical. The
algorithm is implemented by choosing at random between the kernels at each
iteration. (Unbiasedness follows by the Glynn-Rhee trick, which is eminently
well-suited for coupling.) As pointed out from the start of the paper, the
appeal of this unbiased version is that the algorithm can be (embarrassingly)
parallelised since all processors in use return estimators that are i.i.d.~copies
of one another, hence easily merged into a better estimator.}

\section{Approximate Bayesian computation}\label{sec:ABC}

The methods surveyed above share the common feature of exploiting the shape of the
target density, $p(\cdot)$, namely that it is known exactly or known up to a normalising constant 
$p(x)\propto\tilde p(x)$ or yet known as the marginal of another density
$$p(x)=\int_{\mathcal Y} q(x,y)\,\text{d}y\,.$$ 
It may however occur that the density of the target is not numerically
available, in the sense that computing $p(x)$ or $\tilde p(x)$ is not feasible
in a reasonable time or that completing $p(\cdot)$ into $q(\cdot)$ involves a
massive increase in the dimension of the problem. This obviously causes
difficulties in applying, e.g., MCMC methods. A particularly common case occurs
in the Bayesian analysis of intractable likelihoods.

\xval{``What would be a good example of a really simple model that has an
intractable likelihood?" [cross-validated:127180]}{Given an original Normal dataset
$$
x_1,\ldots,x_n\stackrel{\text{iid}}{\sim}\text{N}(\theta,\sigma^2)\,,
$$
the reported data is made of the two-dimensional summary
$$
S(x_1,\ldots,x_n)=(\text{med}(x_1,\ldots,x_n),\text{mad}(x_1,\ldots,x_n))\,,
$$
where $\text{mad}(x_1,\ldots,x_n)$ is the median average deviation of the
sample, which is not sufficient and which does not have a closed form joint density.}

Besides this simple example, there are numerous occurrences of iwell-defined 
likelihoods that cannot be computed, from latent variable models, including 
hidden Markov models, to likelihoods with a missing normalising term depending
on the parameter, including Ising models \citep{potts52} and other non-standard exponential families, 
to densities defined as solutions of differential equations,
via their characteristic function, like $\alpha$-stable distributions
\citep{peters:sisson:fan:2012}, or via
their quantile function like Tukey's $g$-and-$k$ distributions
\citep{haynes:macgillivray:mengersen:1997}. 

A different kind of algorithms is required for handling such situations. They are called `likelihood-free' 
or approximate Bayesian computation (ABC) methods, as they do not require the likelihood function and provide an approximation of the original posterior distribution.

\xval{``What does it mean for an inference
or optimisation method to be `likelihood-free'." [cross-validated:383731]}{
Specifically, likelihood-free methods are a rewording of the ABC
algorithms, where ABC stands for approximate Bayesian computation. This intends
to cover inference methods that do not require the use of a closed-form
likelihood function, but still intend to study a specific statistical model.
They are free from the computational difficulty attached with the likelihood
but not from the model that produces this likelihood. See for instance
the recent handbook by \cite{sisson:fan:beaumont:2019}.}

The basic ABC algorithm is based on the following principle: given a target
posterior proportional to $\pi(\theta) f(x^\text{obs}|\theta)$, when the
likelihood function $f(x^\text{obs}|\theta)$ is not available in closed
form,\footnote{The notation $x^\text{obs}$ is intended to distinguish the
observed sample from simulated versions of this sample.} jointly simulating
$$
\theta^\prime\sim \pi(\theta)\,,z\sim f(z|\theta^\prime)\,,
$$
until the auxiliary variable $z$ is equal to the observed value, $z=x^\text{obs}$
does produce a realisation from the posterior distribution without ever computing a
numerical value of the likelihood function. It only requires that the model associated
with this likelihood can be simulated, which often leads to the model being called a {\em
generative model}.

\xval{``How can we prove that when accepting for $x=x^\text{obs}$
this algorithm, we sample from the true posterior?"
[cross-validated:380076]}{This case is the original version of the algorithm,
as in \cite{rubin:1984} and \cite{tavare:balding:griffith:donnelly:1997}.
Assuming that $$\mathbb{P}_\theta(Z=x^\text{obs})>0$$ the values of $\theta$
that come out of the algorithm are distributed from a distribution with density
proportional to
$$\pi(\theta) \times \mathbb{P}_\theta(Z=x^\text{obs})$$
since the algorithm generates the pair $(\theta,\mathbb{I}_{Z=x^\text{obs}})$ with joint distribution
$$\pi(\theta) \times \mathbb{P}_\theta(Z=x^\text{obs})^{\mathbb{I}_{Z=x^\text{obs}}} \times
\mathbb{P}_\theta(Z\ne x^\text{obs})^{\mathbb{I}_{Z\ne x^\text{obs}}}$$
Conditioning on $\mathbb{I}_{Z=x^\text{obs}}=1$ leads to
$$\theta|\mathbb{I}_{Z=x^\text{obs}}=1 \sim \pi(\theta) \times \mathbb{P}_\theta(Z=x^\text{obs})\Big/\int \pi(\theta) \times \mathbb{P}_\theta(X=x^\text{obs}) \,\text{d}\theta$$
which is the posterior distribution.}

As noted in the above vignette, the principle can only be implemented when
$\mathbb{P}_\theta(Z=x^\text{obs})>0$ and more accurately when the event
$Z=x^\text{obs}$ has a non-negligible chance to occur. This is however rarely
the case in realistic settings, especially when $Z$ is a continuous variable,
and the first implementations \citep{pritchard:seielstad:perez:feldman:1999} 
of the ABC algorithm replaced the constraint of equality $z=x^\text{obs}$ a 
relaxed version,
$$
\varrho(z,x^\text{obs})\le\epsilon
$$
where $\varrho$ is a distance and $\epsilon>0$ is called the tolerance. This
approximation step makes the concept applicable in a wider range of settings
with an intractable, but it also implies that the simulated distribution is
modified from the true posterior into 
$$
\pi(\theta|\varrho(Z,x^\text{obs})<\epsilon)
\propto \pi(\theta)\,\mathbb P_\theta\{\varrho(Z,x^\text{obs})<\epsilon\}\,.
$$
It helps to visualise this alternative posterior distribution as truly conditioning
on the event $\varrho(Z,x^\text{obs})<\epsilon$ rather than $x^\text{obs}$ as it gives
a specific meaning to this distribution and explains the loss in information brought by
the approximation.
 
In many settings, especially with large datasets, looking at a distance between
the raw observed data and the raw simulated data is very inefficient. It is
much more efficient \citep{li:fearnhead:2018,frazier:martin:robert:rousseau:2018} 
to compare informative summaries of the data as the decrease in dimension allows 
for smaller tolerance, a higher signal to noise ratio, and outweighs the potential 
loss in information. A more common implementation of the algorithm is thus

\begin{algorithm}[H]
\caption{\small Likelihood-free (ABC) rejection sampler\label{algo:ABC2}}
\begin{algorithmic}
\footnotesize
\FOR {$i=1$ to $N$}
        \REPEAT
        \STATE generate $\theta'$ from the prior distribution $\pi(\cdot)$
        \STATE generate $z$ from the likelihood $f(\cdot|\theta')$
        \UNTIL {$\rho\{\eta(z),\eta(x^\text{obs})\}\leq \epsilon$}
        \STATE set $\theta_i=\theta'$
\ENDFOR
\end{algorithmic}
\end{algorithm}

\noindent where $\eta(\cdot)$ denotes a (not necessarily sufficient) statistic, 
usually (needlessly) called a summary statistic. While there is a huge literature 
\citep{aeschbacher:beaumont:futschik:2012,fearnhead:prangle:2012,estoup:etal:2012,
blum:etal:2013,sisson:fan:beaumont:2019} on the choice of the summary statistic,
compelling arguments \citep{fearnhead:prangle:2012,li:fearnhead:2018} lead to opt
for summaries of the same dimension as the parameter $\theta$.

While the motivation for simulating from the prior distribution is clear from a
theoretical perspective, given that the probability of accepting in Algorithm
\ref{algo:ABC2} is approximately the intractable likelihood, it is also often
poor in efficiency since the posterior is much more concentrated. Subsequent
versions of ABC have thus aimed at alternative approaches to increase the
efficiency of the method. For instance, the proposal distribution on $\theta$
can be modified towards increase the frequency of $x$'s within the vicinity of $x^\text{obs}$
\citep{marjoram:etal:2003,bortot:coles:sisson:2007,li:fearnhead:2018}. Others
have replaced the indicator function in Algorithm \ref{algo:ABC2} with less
rudimentary estimators of the likelihood
\citep{beaumont:zhang:balding:2002,blum:2010,mengersen:pudlo:robert:2013},
interpreting the tolerance $\epsilon$ as a bandwidth
\citep{li:fearnhead:2018,frazier:martin:robert:rousseau:2018} or a new
component in the inferential framework \citep{ratmann:andrieu:wiuf:richardson:2009}. 

Computational advances have seen MCMC, SMC \citep{beaumont:cornuet:marin:robert:2009} and Gibbs 
\citep{clarte:robert:ryder:stoehr:2019} versions of ABC. For
instance, ABC-MCMC \citep{marjoram:etal:2003} is based on the property that the
Markov chain $(\theta^{(t)})$ created via the transition function
\[
\theta^{(t+1)}= \begin{cases}
        \theta^\prime\sim K_\omega(\theta^\prime|\theta^{(t)}) &\text{if }z\sim f(z|\theta^\prime)\text{ is such that }z=x^\text{obs}\\
                &\text{ and }u\sim\mathcal{U}(0,1) \le \frac{\pi(\theta^\prime)
                K_\omega(\theta^{(t)}|\theta^\prime)}{\pi(\theta^{(t)}) K_\omega(\theta^\prime|\theta^{(t)})}\,,\\
        \theta^{(t)} &\text{otherwise,}
                \end{cases}
\]
enjoys the posterior $\pi(\theta|x^\text{obs})$ as its stationary distribution.
The corresponding algorithm is then

\begin{algorithm}[H]
\caption{\small Likelihood-free MCMC sampler\label{algo:ABCMC}}
\begin{algorithmic}
\footnotesize
\STATE Use Algorithm \ref{algo:ABC2} to get $(\theta^{(0)},z^{(0)})$
\FOR {$t=1$ to $N$}
\STATE Generate $\theta'$ from $K_\omega\left(\cdot|\theta^{(t-1)}\right)$,
\STATE Generate $z'$ from the likelihood $f(\cdot|\theta')$,
\STATE Generate $u$ from $\mathcal{U}_{[0,1]}$,
\IF {$u \leq \frac{\pi(\theta')K_\omega(\theta^{(t-1)}|\theta')}{\pi(\theta^{(t-1)}K_\omega(\theta'|\theta^{(t-1)})}
\mathbb{I}_{\varrho(\eta(z'),\eta(x^\text{obs}))\le\epsilon}$}
\STATE set $(\theta^{(t)},z^{(t)})=(\theta',z')$
\ELSE
\STATE $(\theta^{(t)},z^{(t)}))=(\theta^{(t-1)},z^{(t-1)})$,
\ENDIF
\ENDFOR
\end{algorithmic}
\end{algorithm}

The choice of summary statistics in ABC method is paramount for the efficiency of the
approximation and nowhere more than for model choice. Since the Bayes factor is given by
$$
B_{12}(x^\text{obs}) = \frac{\text{Pr}(M_1|x^\text{obs})}{\text{Pr}(M_2|
x^\text{obs})}\Big/\frac{\text{Pr}(M_1)}{\text{Pr}(M_2)}
$$
the ratio of frequencies of simulations from $M_1$ and $M_2$ that are accepted need
be divided by the prior probabilities of $M_1$ and $M_2$ if these reflect the
number of times each model is simulated. Apart from this, the approximation is valid.
Using inappropriate summary statistics in this setting  has been pointed out in
\cite{didelot:everitt:johansen:lawson:2011},
\cite{robert:cornuet:marin:pillai:2011} and \cite{marin:pillai:robert:rousseau:2011}.

A special instance of (almost) intractable is the setting of ``Big Data" problems where the
size of the data makes computing the likelihood quite expensive. In such cases, ABC can be seen
as a convenient approach to scalable Monte Carlo.

\begin{rema}
\footnotesize
\begin{quote}\begin{em}
What difference does it make working with a big or small dataset in ABC? [cross-validated:424712]
\end{em}\end{quote}

It all depends on the structure of the dataset and the complexity of the model behind. In some settings the size of the data may be the reason for conducting an ABC inference as the likelihood takes too much time to compute. But there is no generic answer to the question since in the ultimate case when there exists a sufficient statistic of fixed dimension size does not matter (and of course ABC is unlikely to be needed).

\begin{quote}\begin{em}
Do we get any computational benefits by reducing a very big dataset
when doing inference using ABC methods?
\end{em}\end{quote}

In most settings, ABC proceeds through a set of summary statistics that are of a much smaller dimension than the data. In that sense they are independent of the size of the data, except that to simulate values of the summaries, most models require simulations of the entire dataset first. Unless a proxy model is used as in synthetic likelihood.

\begin{quote}\begin{em}
...the rejection criterion in ABC is related to how well we
approximate the full likelihood of the dataset which is typically
captured in some low-dimensional summary statistics vector.
\end{em}\end{quote}

You have to realise that the rejection is relative to the distribution of the distances between the observed and the simulated summaries [simulated under the prior predictive], rather than absolute. In other words, there is no predetermined value for the tolerance. This comes in addition to the assessment being based on an insufficient statistics rather than the full data. This means that, for a given computing budget, the true likelihood of an accepted parameter may be quite low. 
\end{rema}

\section{Further reading}

There are many reviews and retrospective on the Markov Chain Monte Carlo
methods, not only in statistics, but also in physics, econometrics and several
other fields, most of which provide different perspectives on the topic. For
instance, \cite{dunson:johndrow:2020} recently wrote a celebration of Hastings'
1970 paper in Biometrika, where they cover adaptive Metropolis
\citep{haario:sacksman:tamminen:1999,roberts:rosenthal:2005}, the importance of
gradient based versions toward universal algorithms
\citep{roberts:tweedie:1995,neal:2003}, discussing the advantages of HMC over
Langevin versions. They also recall the significant step represented by Green's
(\citeyear{green:1995}) reversible jump algorithm for multimodal and
multidimensional targets, as well as tempering
\citep{woodard2009sufficient,miasojedow2013adaptive}. They further cover
intractable likelihood cases within MCMC (rather than ABC), with the use of
auxiliary variables \citep{moeller:pettitt:reeves:2006,friel:pettitt:2008} and
pseudo-marginal MCMC \citep{andrieu2009pseudo,andrieu:vihola:2014}. They
naturally insist upon the need to handle huge datasets, high-dimension
parameter spaces, and other scalability issues, with links to unadjusted Langevin schemes
\citep{welling2011bayesian,bardenet:etal:2014,durmus:moulines:2017}. Similarly,
\cite{dunson:johndrow:2020} discuss recent developments towards parallel MCMC
and see non-reversible schemes such as PDMP as highly promising, with a
concluding section on the challenges of automating and robustifying much
further the said procedures, if only to reach a wider range of applications.
Other directions that are clearly still relevant after decades of development include
convergence assessment, e.g. the comparison and aggregation of various
approximation schemes, since this is a fairly common request from users, 
recycling schemes, like Rao-Blackwellisation \citep{gelfand:smith:1990,
casella:robert:1996} and other post-processing improvements that address the massive
waste of simulation in most method, the potential for mutual gains between machine-learning tools
and MCMC refinements, as well as the theoretical difficulties presented by approximations such as
synthetic likelihood \citep{wood:2010}, indirect inference \citep{drovandi:pettitt:faddy:2011} and
incompatible conditionals \citep{plummer:2015,jacob:etal:2017,clarte:robert:ryder:stoehr:2019}.

\small
\renewcommand\bibsection{\section{\bibname}}
%\bibliographystyle{apalike}
%\bibliography{reference.bib}
\hyphenation{Post-Script Sprin-ger}

\end{document}